\documentclass[11pt]{article}
\usepackage[centertags]{amsmath}
\usepackage[square, comma, sort&compress,numbers]{natbib}
\usepackage{array,multirow}
\numberwithin{equation}{section}
\usepackage{amssymb}
\usepackage{comment}
\usepackage{graphicx}
\usepackage{float}
\usepackage{color}

\usepackage{epsfig}
\usepackage{tikz,pgf}
\usetikzlibrary{shapes}
\usetikzlibrary{calc}
\usetikzlibrary{decorations.pathmorphing}
\usetikzlibrary{decorations.pathreplacing,shapes.misc}
\usetikzlibrary{positioning}
\usetikzlibrary{arrows}
\usetikzlibrary{decorations.markings}
\tikzset{snake it/.style={decorate, decoration=snake}}

\usepackage{mathrsfs}

\def\be{\begin{equation}}
\def\ee{\end{equation}}

\newcommand{\bref}[1]{\textbf{\ref{#1}}}

\def\tr{{\rm Tr}}

\def\cW{\mathcal{W}}

\newcommand{\CC}{\mathbb{C}}
\newcommand{\RR}{\mathbb{R}}

\newcommand{\Tr}{\mathrm{Tr}}

\newcommand{\dd}{\mathrm{d}}
\newcommand{\pd}{\partial}

\newcommand{\Res}{\mathrm{Res}}

\newcommand{\OO}{\mathcal{O}}

\newcommand{\ba}{\begin{equation}\begin{aligned}}
\newcommand{\ea}{\end{aligned}\end{equation}}

\usepackage{jheppub}
\makeatletter
\def\@fpheader{\vspace{-.1cm}}
\makeatother

\title{Open minimal strings and open Gelfand-Dickey hierarchies}

\author[a,b]{Konstantin\ Aleshkin,}
\author[c,d,e\,\dagger]{\;Vladimir\ Belavin}
\note[$\dagger$]{Weston Visiting Professorship at Weizmann Institute.}

\affiliation[a]{L.D. Landau Institute for Theoretical Physics,\\
 Akademika Semenova av. 1-A, 142432 Chernogolovka,  Moscow region, Russia}
\affiliation[b]{International School of Advanced Studies (SISSA),\\
 via Bonomea 265, 34136 Trieste, Italy}
\affiliation[c]{I.E. Tamm Department of Theoretical Physics, P.N. Lebedev Physical
Institute,\\ Leninsky ave. 53, 119991 Moscow, Russia}
\affiliation[d]{Department of Quantum Physics,
Institute for Information Transmission Problems, \\
Bolshoy Karetny per. 19, 127994 Moscow, Russia}
\affiliation[e]{Department of Particle Physics and Astrophysics, Weizmann Institute of Science,\\
Rehovot 7610001, Israel}
\emailAdd{kaleshkin@sissa.it}
\emailAdd{belavin@lpi.ru}

\abstract{
We study the connection between minimal Liouville string theory and generalized open KdV hierarchies. We are interested in generalizing Douglas string equation formalism to the open topology case. We show that combining the results of the closed topology, based on the Frobenius manifold structure and resonance transformations,
with the appropriate open case modification, which requires the insertion of macroscopic loop operators, we reproduce the well-known result for the expectation
value of a bulk operator for the FZZT brane coupled to the general  $(q,p)$ minimal model. The  matching of the results of the two setups gives new evidence of the connection between minimal Liouville gravity and the theory of Topological Gravity.
}


\begin{document}
\maketitle
\flushbottom

%=======================================
\section{Introduction} \label{section:1}
%=======================================

Minimal string theory is a self-consistent solvable model,
which represents a valuable source of insight for investigating string dynamics. 
It is based on $(q,p)$ minimal CFT models, appropriately coupled to 2D Liouville gravity.
There are two approaches to minimal strings, or equivalently to minimal Liouville gravity (MLG): the direct  approach, based on the worldsheet treatment~\cite{Polyakov1} and the  dual approach, originating from the matrix models, and leading ultimately to the Douglas string equation formalism~\cite{Douglas:1989dd}.

During last few years there has been a new progress in the dual approach,  which arose due to the connection~\cite{Belavin:2013nba} between the Douglas string equation formalism and the structure of a certain class of Frobenius manifolds (FM). 
\footnote{The role of Frobenius manifolds has been initially established~\cite{Dubrovin:1992dz} in the context of topological field theories.} The new development~\cite{Belavin:2013nba,Belavin:2014cua, Belavin:2014hsa,Belavin:2014xya,Belavin:2015ffa}  allowed to relate the generating function of MLG correlation numbers (or minimal string scattering amplitudes) in the spherical topology\footnote{For some further progress in MLG on a sphere, see~\cite{Tarnopolsky:2009ec,Belavin:2010bs,Belavin:2010sr,Spodyneiko:2015xpa}.} with the dispersionless tau-function of the corresponding hierarchy, for which a simple closed representation has been found, using the underlying  FM structure. It turns out that a crucial role in this construction plays a so-called resonance transformation~\cite{MSS,Belavin:2008kv} from  KdV times  to Liouville coupling constants, which was found explicitly in the general $(q,p)$ case in~\cite{Belavin:2014hsa} up to a certain order.

In this paper we continue our study~\cite{Aleshkin:2017yty} (see also \cite{Bawane:2018ejq,Bawane:2018zwa,Muraki:2018rqv}, focused on $(2,2k+1)$  Lee-Yang series) of the boundary version of minimal Liouville gravity (BMLG). 
In the worldsheet approach, constructing correlation numbers, apart from the moduli integration problems~\cite{Aleshkin:2016snp},  needs a set of new ingredients~\cite{Cardy:1984bb,Cardy:1986gw,Cardy:1991tv,Runkel:1998he,Fateev:2000ik,Ponsot:2001ng}, connected with the boundary,  and represents quite a non-trivial problem. The development of the  
dual approach to BMLG hence represents a natural task (see~\cite{MSS,Kostov:2003uh, Bourgine:2009zt,Bourgine:2008pg,Kostov:2002uq, Jacobsen:2006bn,Ishiki:2010wb,Martinec:1991ht,Hosomichi:2008th} and references therein, for some relevant results within the KdV frame). 
In the case of Lee-Yang series, the boundary version of the dual partition function~\cite{MSS} can be rather easily translated~\cite{Belavin:2010ba} to the Liouville frame, taking into account the resonance transformation. From the Frobenius manifolds perspective,  the reason for this simplification  is that the FM, related to Lee-Yang series, is trivial, i.e., one-dimensional, and the way to define a matrix model,  corresponding to $(2,2k+1)$ BMLG model, is unique. In general $(q,p)$ case~\cite{Belavin:2013nba,Belavin:2014hsa}, multi-dimensional  Frobenius manifolds arise  and higher Gelfand-Dickey hierarchies appear, which makes the correspondence less straightforward.

In this work we study the bulk correlations on a disk. 
The goal is to clarify how to combine the open topology treatment (e.g. macroscopic loops~\cite{MSS}) with the formalism, based on the Frobenius manifold structure. Important ingredients of topological gravity are
Virasoro constraints~\cite{Dijkgraaf:1990rs}, which are equivalent to the KdV equations~\cite{Douglas:1989dd} in $(2,2p+1)$ series, complemented by the string
equation. For the boundary partition function one can write so-called
open Virasoro constraints~\cite{Buryak:2014dta} which generalize classical ones
and fix the partition function. We describe the consequences of these constraints for the minimal Liouville gravity.

In the general $(q,p)$ case, the Virasoro constraints are particular case
of $\cW$-constraints, which can be realized using bosonization in terms of a set of
twisted free boson fields.
This allows to define an open string partition function as a vertex exponential  operator, constructed form the  bosons,  and gives simple expression for the generating function on a disk, clarifying the connection with the open $\cW$-constraints. We check the expression for one-point  bulk correlators, and reproduce the results of the worldsheet treatment for FZZT brane~\cite{Fateev:2000ik,Teschner:2000md}. The matching of the results confirms the universality of the resonance transformation and its explicit form  found on a sphere in~\cite{Belavin:2015ffa}.   
We also compare this approach to the formula for bulk one point
functions on a disk used in~\cite{Aleshkin:2017yty}. It turns out that the formulain~\cite{Aleshkin:2017yty} is a particular
case of the macroscopic loop described in the present paper.
Moreover, the equality of two formulas is a version of mirror symmetry formula
for $q-$spin intersection numbers and $A_{q-1}$ Landau-Ginzburg models.

The paper is organized as follows. In Section~\bref{section:prelim} we briefly remind the definition of MLG, describe the physical fields, and introduce the boundary one
point correlation numbers. In Section~\bref{section:Lee-Yang} we consider one matrix model and minimal gravity of type $(2,2k+1)$. 
In Section~\bref{section:general} we consider the general $(q,p)$ case.
In Section~\bref{section:previous} we articulate the new points, with respect to our previous consideration~\cite{Aleshkin:2017yty}, which at that time was essentially limited by the unitary series $(q,q+1)$.  In Section~\bref{section:concl} we conclude, discussing some open questions. Some technical details are collected in Appendix~\bref{sec:appendix}.

%=======================================
\section{Minimal Liouville Gravity}
\label{section:prelim}
%=======================================
In this section we briefly recall the worldsheet formulation of (bosonic) MLG (for more details, see, e.g., \cite{Seiberg:2003nm}), in order to introduce the bulk correlators, that
we are interested in this paper.

\subsection{Worldsheet formulation}

\paragraph{MLG definition.} Minimal Liouville gravity is non-critical string theory model~\cite{Polyakov1, KPZ}, obtained by combining
bosonic $(q,p) $ minimal model~\cite{BPZ} (with $p>q\geq2$ being relatively prime integers), with ghosts and Liouville field theory~\cite{ LFT}, described by scalar boson field $\phi(z)$, arising due to the
conformal anomaly effect. The MLG partition function splits into three parts
\be
Z_{\text{MLG}} = Z_{\text{MM}} \cdot Z_{\text{Liouville}} \cdot Z_{\text{gh}}\;, 
\ee
where all three sectors obey conformal symmetry and the zero total central charge condition,
$c_{\text{MM}}+c_{\text{Liouville}} -26=0$, which follows from the Weyl invariance of the string action and ensures 
BRST invariance of MLG theory. This condition constraints the Liouville coupling constant to be $b=\sqrt{q/p}$.

In this paper we are interested in the physical operators, or BRST cohomologies $\OO_{m,n}$, constructed by dressing the minimal model primaries $\Phi_{m,n}(z)$  (where $1\leq m\leq q-1$ and  $1\leq n\leq p-1$, modulo Kac symmetry, implemented below by restricting $p m-q n> 0$)
with the Liouville {\it non-degenerate} primary exponential fields $V_a(z)\equiv :e^{2a \phi(z)}:$
\be\label{tachyon}
\OO_{m,n} = \int \dd^2 z \Phi_{m,n}(z) V_{a_{m,-n}} (z)\;,
\ee
where $a_{m,-n}$ is the solution of  the dimensional constraint 
$\Delta^{\text{MM}}_{m,n}+\Delta^{\text{Liouville}}(a)=1$, in the standard Liouville parametrization~\cite{LFT}. We note that apart from these {\it tachyon} physical operators~\eqref{tachyon}, there exist an important class of {\it ground ring} operators,
constructed from {\it degenerate} Liouville primaries, which is relevant in particular for constructing  multi-point tachyon correlation numbers in the worldsheet approach~\cite{Belavin:2006ex,Belavin:2009cb}, and which will not be considered in this paper.

\paragraph{Generating function.} One can pack multi-point bulk correlators $\langle \OO_{m_1,n_1}...\OO_{m_N,n_N}\rangle_{g}$
in a generating function $Z_{g}(\lambda)$,  where $g$ is a genus and $\lambda=\{\lambda_{m,n}\}$ is a set of (MLG) Liouville coupling constants,~\footnote{We do not
discuss here subtleties of the gauge fixing and defining  spherical 1- and 2-point functions.}
\be\label{gen-fun}
Z_{g}(\lambda) = \langle \exp\left(\sum \lambda_{m,n} \OO_{m,n} \right) \rangle_g\;.
\ee
Our discussion here is limited to a sphere, $g=0$,  or a disk worldsheet topology, relevant for the planar limit, so that in what follows we omit the subscript $g$, meaning either spherical or disk topology. 
The coupling $\lambda_{1,-1}=\mu$  is the Liouville bulk cosmological constant, which corresponds to the operator $e^{2b\phi(x)}$, measuring the area of the worldsheet Riemann surface.
The generating function~\eqref{gen-fun} has simple scaling properties, being quasi-homogeneous function of the cosmological constant $\mu$, which allows to assign~\cite{KPZ}  the gravitational dimensions $\delta_{m,n}$ to the physical fields $\OO_{m,n}$, and to $Z(\lambda)$.
This quasi-homogeneity property together with the conformal selection rules for admissible correlation functions are important constraints for constructing the dual representation of the generating function via the resonance transformation~\cite{Belavin:2014hsa}, which will be described in Section~\bref{section:general}.

\subsection{Bulk correlators on a disk}

In the open case one needs to appropriately define the boundary conditions
both in the matter minimal model and in the Liouville sector. The former are conformally invariant 
Cardy states, labeled by two integers $(r,s)\in Kac(q,p)$, that is
$1\le r\le q,\; 1\le s\le p$ with identification $t_{r,s} \sim t_{q-r,p-s}$,
and the later can be either FZZT brane~\cite{Fateev:2000ik,Teschner:2000md} or ZZ  brane~\cite{Zamolodchikov:2001ah}. We will focus here on the FZZT brane, 
which is specified by the value of the boundary Liouville cosmolgical
constant $\mu_B$. We recall that the boundary Liouville interaction has the form 
$\mu_B\int_{\partial \Gamma} e^{b\phi(x)} \dd \xi$, where $\partial \Gamma$ is the boundary of the disk and the integral can be interpreted as the length $l$ of the boundary,  so that the Laplace transform of the boundary partition function with respect to $\mu_B$ gives the dependence on 
the boundary length $l$.
  
Denoting by $|B_{r,s}(\mu_B)\rangle$ the open minimal string boundary states, which is obtained by tensoring together the Cardi matter states and the Liouville FZZT sates, we write  the generating function of the  bulk multi-point correlators on a disk
\be
\langle \OO_{m_1,n_1}...\OO_{m_N,n_N}| B_{r,s}(\mu_B)\rangle\equiv 
\langle \OO_{m_1,n_1}...\OO_{m_N,n_N}\rangle_D
\ee
as
\be\label{gen-fun-disk}
Z_{D}(\lambda,\mu_B) = \langle \exp\left(\sum \lambda_{m,n} \OO_{m,n} \right)|B_{r,s}(\mu_B)\rangle
\equiv\langle \exp\left(\sum \lambda_{m,n} \OO_{m,n} \right)\rangle_D\;.
\ee
Our goal is to study this generating function from the perspective of the dual approach.  
Note that in the boundary case  the correlators depend on the scale invariant ratio $\mu/\mu_B^2$, so that
\be
\langle \OO_{m_1,n_1}...\OO_{m_N,n_N}| B_{r,s}(\mu_B)\rangle \sim \mu^{(Q-2 \sum_i \delta_{m_i,n_i})/2b} F\left(\frac{\mu_B^2}{\mu}\right)\;,
\ee
where $Q=b+1/b$. Since the Cardy states with $(k,l)=(1,1)$ form a basis in the space of FZZT branes (see~\cite{Seiberg:2003nm}),
we will focus on $\mu_B$ (or $l$) dependence of the correlators.

One of the simplest amplitude to compute in such a theory is a bulk insertion
one-point function on a disk.  Formula for such an amplitude was computed 
in~\cite{Fateev:2000ik} with the result (as a function of boundary length 
instead of $\mu_B)$:
\be \label{FZZLoop}
\langle \OO_{m,n} \rangle_D \simeq \kappa^{\nu_{m,n}} K_{\nu_{m,n}} (\kappa l)
\ee
and $\kappa^2 = \mu/\sin(\pi b^2)$. We do not specify the constants, because
in this paper we compute just the functional dependence on $\mu$ and $l$.\\

In the rest of the paper we will focus on matrix models/integrable systems 
approach and comparison with the Liouville approach. Namely, we will show how
the FZZ formula~\eqref{FZZLoop} appears from another formulation of
the loops in (q,p) model and thus, how the integrable systems
equation enter the MLG picture.

%%%%%%%%%%%%%%%%%%%%%%%%%%%%%%%%%%%%%%%%%%%%%%%%%%%%%%%%%%%%%%%%%%%%%%%%%%%%%%

%=======================================
\section{Virasoro constraints and  (2,2k+1) minimal gravity}
\label{section:Lee-Yang}
%=======================================

In this section, we recall the matrix model approach to the Lee-Yang series of MLG, 
we describe the Virasoro constraints in the closed and then in the boundary case. We follow mostly the recent review of the subject in~\cite{Dijkgraaf:2018vnm}.  

Two basic observables in the matrix models are the trace operators $\tr M^n$ and the resolvent $\Tr (x - M)^{-1}$,    
 the goal of this section is to show that a modified version  of the resolvent can be interpreted as a chiral  free  boson field $\phi(x)$, defined in terms of creation and annihilation operators acting on the space of functions of the matrix model parameters.  The free boson $\phi(x)$  naturally obeys the (half-)Virasoro symmetry with the central charge $c=1$. This fact leads to the  Virasoro constraints in the closed topology case and gives a clear CFT interpretation of the boundary insertion as an insertion of a primary vertex constructed from the boson. One of the conceptual points here is that while for finite size matrix $M$ we are dealing with the standard chiral boson, defined on the $x$-plane, in the double scaling limit relevant for MLG we are lead to consider a boson on a hyper-elliptic spectral curve, which is double cover of the $x$-plane. This consideration will be used further in order to consider the general  $(q,p)$ case,
where instead of one boson $q-1$ twisted bosons arise.         

\subsection{One-matrix model and Virasoro constraints}

The perturbative expansion of matrix models (MM) in terms of  ribbon Feynman diagrams gives  an interpretation of MM as a discrete version of 2D quantum gravity~\cite{Douglas:1989dd,Kazakov:1989bc,Brezin:1990rb}.

The $(2,2p+1)$ MLG model is described by the double scaling limit $N\rightarrow\infty$ around the $(p+1)$th critical point of a one-matrix model, which consists of a random $N\times N$ hermitian matrix $M$, with the partition function  
\be \label{pf}
Z(t) = \frac{1}{\text{vol}(U(N))}\int_{N\times N} \dd M\cdot e^{-\frac{1}{g} \Tr W(M)} = \int_{\RR^N} \dd^N \lambda\, \Delta(\lambda)^2 e^{-\frac{1}{g}\sum_{j=1}^N W(\lambda_j)}\;,
\ee
where  $t=\{t_n\}$, the potential $W(M) = \sum_{n} t_n M^n$, and $g$ is the string coupling constant. The model obeys $U(N)$ gauge invariance, $M\rightarrow U M U^{-1}$,  and the RHS expression in~\eqref{pf}  is the result of taking this symmetry into account, with $\lambda$ being the vector of eigenvalues of $M$, and $\Delta(\lambda)$ being the Vandermonde determinant. The correlators are given by the derivatives of $\log Z(t)$ with respect to the coupling parameters $t_n$. 

For any $N$ the matrix model enjoys (half-)Virasoro symmetry.  In order to derive the Virasoro constraints,  we define
\be \label{bos1}
J(x) =   W'(x)/2 - g \Tr \frac{1}{x - M}\;,  
\ee 
where the second term is the matrix resolvent $\Tr (x - M)^{-1}= \sum_k (x-\lambda_k)^{-1}$.
Then, one can check that 
\be \label{ward1}
\langle J(x)^2 \rangle = \langle W'(x)^2/4 - \sum_k \frac{W'(x) - W'(\lambda_k)}{x - \lambda_k} \rangle\;,
\ee
where the angle brackets denote averaging with the measure defined in~\eqref{pf}, and the expression on the right hand side is  easily seen to be regular at $x=0$. 

Inside the matrix integral, $J(x)$ can be written as a differential operator,
\be \label{bos1}
J(x) = \sum_{n>0} n t_n x^{n-1}/2 - g\sum_{n\ge 0} \frac{\partial}{\partial t_n} x^{-n-1}\;,  
\ee
which up to normalization is the current of the chiral free boson $\phi(x)$, 
\be\label{J-phi}
J(x) =\frac{g}{\sqrt{2}}\pd \phi(x)\;.
\ee
Hence, one can consider an energy momentum tensor of this boson, $T(x) = :\pd \phi(x)^2:$, where dots stand the normal ordering, and the equation~\eqref{ward1} takes the form 
\be
\langle T(x)\rangle = reg\;, \; \text{ or } T(x)Z(t) = 0\;, \; x \to 0\;,
\ee
where in the second equality we consider $T(x)$ as a differential operator acting
on functions of $t$.
If we expand $T(x) = \sum_n L_n x^{-n-2},$ then the latter equation can be written in terms of Virasoro constraints:
\be
L_n Z(t) = 0, \; n \ge -1\;,
\ee
where $L_n$ are found from the definition of $T(x)$:
\ba
&L_{-1} = \sum_{k\ge 1} k t_k \frac{\pd}{\pd t_{k-1}}, \\
&L_n = \sum_k k t_k \frac{\pd}{\pd t_{k+n}} + g^2 \sum_{i+j=n} \frac{\pd^2}
{\pd t_i \pd t_j}, \quad n \ge 0
\ea
and by construction satisfy Virasoro algebra with $c=1$.

In the double scaling limit the Virasoro structure also persists with the deference that
the boson becomes twisted, that is it acquires half-integer modes in the expansion
\be \label{twistedboson}
\frac12 g \pd \phi(x) = x^{1/2} - \sum_{n\ge0}(n+1/2)t_n x^{n-1/2}-
\frac14 g^2\sum_{n\ge0}\frac{\pd}{\pd t_n} x^{-n-3/2}\;.
\ee
Here now $t_n$ stand for new (re-scaled) KdV couplings, which are some functions of the ``bare'' couplings of the underlying matrix model~\eqref{pf}. 
The derivatives $\pd/\pd_{t_k}$ are interpreted as insertions of operators
$\OO_k$ in the correlation function.
In order to motivate this change we note that in the
semiclassical limit the eq.~\eqref{ward1}  becomes
\be \label{speccurve}
y^2 = P(x)\;,
\ee
where $y:= \lim_{N\to\infty}\langle J(x) \rangle$, and
$P(x)$ is a polynomial, which arises from the RHS of~\eqref{ward1}. This can be interpreted as an equation for
a so-called spectral curve. The boson $\phi(x)$ is then defined on this curve rather
then on the $x$-plane (for more details, see e.g.~\cite{Dijkgraaf:2018vnm}).

In the double scaling  limit the Virasoro constraints, which arise for the twisted free boson~\eqref{twistedboson}, become:
\ba \label{vir1}
&L_{-1} = -\frac{\pd}{\pd t_0}+\sum_{k\ge 1} (k+1/2) t_k \frac{\pd}{\pd t_{k-1}}+
\frac{1}{2g^2}t_0^2\;, \\
&L_{0} = -\frac{\pd}{\pd t_1}+\sum_{k\ge 0} (k+1/2) t_k \frac{\pd}{\pd t_{k}}+
\frac{1}{16}\;, \\
&L_{n} = -\frac{\pd}{\pd t_{n+1}}+\sum_{k\ge 0} (k+1/2) t_k \frac{\pd}{\pd t_{k+n}}+
\frac{1}{8}g^2 \sum_{i+j=n-1} \frac{\pd^2}{\pd t_i \pd t_j}\;,\qquad n>0\;. \\
\ea

Finally we note, that any function of KdV parameters, which is annihilated  by all these operators, is uniquely defined and 
represents in fact a (square root of a) tau-function of a KdV hierarchy~\cite{Makeenko:1990in}.

\subsection{Open Liouvulle problem}

It is well-known that the correlations on a surfaces with boundaries can be considered 
by adding loops to the Feynman  diagram expansion of the matrix integral~\eqref{pf}. 
One way to add loops is using additional vector degrees of freedom, contributing to the matrix integral as
\be
\int \dd \Psi \dd \bar{\Psi} \cdot e^{-z \bar{\Psi} \Psi+\bar{\Psi}^T \cdot M 
\cdot \Psi}\;.
\ee
The only modification of the partition function, after integrating out these variables, is the insertion of an additional  determinant in
 the closed matrix integral
\be\label{Z-open}
Z_{open}(t) = \frac{1}{\text{vol}(U(N))} \int \dd M \, \det(z-M)^{N_b} e^{-\frac{1}{g}\Tr W(M)}\;,
\ee
where we denoted by  $N_b$ the number-of-boundaries counting parameter. Looking at the new element $\det(z-M)^{N_b}$ from the free boson perspective,  it is not difficult to
see, that the boson $\phi(x)$  can be defined now  according to~\eqref{J-phi},
 and 
\be \label{boson2}
J(x) = \sum_{n>0} n t_n x^{n-1}/2 - g^2\sum_{n\ge 0} \pd_n x^{-n-1}
+ \frac {g N_b}{2(z-x)}\;,
\ee
which obeys the Virasoro symmetry, with the stress tensor $T(x) = J(x)^2/g^2$.  
It is convenient to introduce an extra factor $e^{-N_b W(z)/(2g)}$ in the matrix integral, which is trivial because it does not depend on the matrix variables,
\be\label{Z-open-2}
Z(t,z) := \frac{1}{\text{vol}(U(N))} \int \dd M \, \det(z-M)^{N_b} \, e^{-\frac{N_b}{2g} W(z)}\, e^{-\frac{1}{g}\Tr W(M)}\;.
\ee
In this setting $J(x)$ as a differential operator 
is given by the same formula~\eqref{bos1} as in the closed string case due to the
factor $e^{-N_b W(z)/(2g)}$.

The new point is that the refined boundary partition function~\eqref{Z-open-2} is equivalent to
\be
\langle \det(z-M)^{N_b} e^{-N_b W(z)/2g} \rangle = \langle V(z) \rangle\;,
\ee
 where  $V(z) = :e^{-N_b\phi(z)/\sqrt{2}}:$ is the primary vertex operator, constructed form the new boson. 

Repeating computation of the closed case we get:
\be \label{wardopen}
\langle T(x) \cdot V(z) \rangle 
= \frac{N_b^2}{4(x-z)^2} V(z) + \frac{1}{x-z}\pd_z V(z) +P(x)\;,
\ee
where $P(x)$ is regular at $x=0$. We can now expand all the quantities in
series in powers of $x$ and consider the terms at negative powers:
\be
(L_n^c+L_n^o) \langle V(z) \rangle = 0, \; \quad n\ge -1\;.
\ee
Here $L_n^c$ come from the expansion of $T(x)$ and are the same
as in the closed case~\eqref{vir1} and $L_n^o$ come from the right hand side
of~\eqref{wardopen}:
\be
L_n^o = -z^{n+1} \frac{\pd}{\pd z} - \frac{N_b^2}4(n+1) z^n\;.
\ee

Now we construct the loop operator, which corresponds to a surface
with one boundary component. To this end we take the first coefficient
in the  series expansion in $N_b$,
% of the full open partition function
\ba \label{loopFormula2}
&w(z)  = \frac{\pd}{\pd N_b} \langle \det(z-M)^{N_b} e^{-\frac{N_b}{2g} W(z)} \rangle|_{N_b=0}
= \\ =&
\langle \left(-\frac1{2g} W(z) + \tr \frac{1}{z-M}\right) e^{-\frac{N_b}{2g} W(z)} \rangle 
\sim \langle \phi(z) \rangle\;.
\ea
In order to compare with the result of the direct minimal Liouville gravity approach we introduce the Laplace transform
of $w(z)$, according to $w(l)=
l^{-1}\int_{0} \dd z \, e^{-lz} \, w(z)$. The singular part of this operator 
is given by~\eqref{boson2},
and the loop~\eqref{loopFormula2}  takes the form of the following differential operator
\be \label{loop2}
w(l) = \sum_{k=0}^{\infty} \frac{l^{k+1/2}}{\Gamma(k+1/2+1)} \OO_{k}\;.
\ee

\section{$\cW$-constraints and loop operator in general (q,p)-case}
\label{section:general}

With the  insight from the one-matrix model now we consider the general situation. Our goal will be to explicitly construct the generalization of the open generating function with the loop operator~\eqref{loop2}  and to check it against  the
results of the worldsheet approach. 
   
\subsection{From twisted bosons to loop operator}

General $(q,p)$ models have several matrix model descriptions, some of which like 
conventional multi-matrix model are appropriate for orthogonal polynomials method and for the double scaling limit consideration.
It turns out, however, that the $\cW$-symmetry which is present in the general $(q,p)$ case
is not manifest in this setup. What we use here is the integrable systems approach, which can be also obtained from the conformal matrix models approach~\cite{Kharchev:1992iv}. In this setting various quantities of the theory are 
expressed by analogy with the one-matrix case. The results of this approach can be briefly
formulated as follows (see, e.g.,~\cite{Johnson:1993vk}).

We consider $q$ twisted bosons $\phi_l(x)$, with $l=1,...,q$:
\be
\pd \phi_l(x) = \sum_{k=-\infty}^{\infty} \alpha_{k+l/q} x^{-(k+l/q+1)}\;,
\ee
where the modes
\be
\alpha_{k+l/q} = g \frac{\pd}{\pd t_{l,k}} \text{  and  } \alpha_{-k-l/q} = 
\frac1{g}(k+l/q) t_{l,k}\;.
\ee
The energy-momentum tensor is
\be
T(x) = \sum_{r=1}^{q-1} :\frac12\pd\phi_r \pd\phi_{q-r}:+\frac{q^2-1}{24qz^2}=
\sum_{n=-\infty}^{\infty} L_n x^{n-2}.
\ee
The system obeys an extended $\cW_{q-1}$ symmetry,  and the other $\cW$-currents, $W^{(n)}(z)$, 
may be constructed explicitly using the standard bosonization methods.
The closed string partition function $Z(t)$ of the $q$-th model is uniquely
defined by the condition that it is annihilated by all the $W^{(n)}$-currents~\cite{Goeree:1990xq}.
More precisely,
\be
W^{(n)}_k Z(t) = 0\;, \quad n\ge2\;, \quad k\ge 1-n\;,
\ee
where $W^{(2)}_k$ is $L_k$ and the correlators are given as usual as
\be
\langle \OO_{\alpha_1,k_1} \cdot \ldots \cdot \OO_{\alpha_n, k_n} \rangle :=
\frac{\pd}{\pd t_{\alpha_1, k_1}} \ldots \frac{\pd}{\pd t_{\alpha_n, k_n}}
\log Z(t)\;.
\ee

For later purposes, we write explicitly the string and the dilation equations, $L_{-1}Z(t)=0$ and $L_{0}Z(t)=0$, with the generators:
\ba \label{strdil}
&L_{-1} = \sum_{\alpha,k} (\alpha/q+k) t_{\alpha,k}
\frac{\pd}{\pd t_{\alpha,k-1}} + 
\frac1{2g^2} \sum_{\beta} \beta(q-\beta)/q^2 t_{\beta,0}t_{q-\beta,0} \;,\\
&L_{0} = \sum_{\alpha,k} (\alpha/q+k) t_{\alpha,k}
\frac{\pd}{\pd t_{\alpha,k}} + 
\frac{q^2-1}{24q}\;.
\ea
Another equivalent description of the system is based on the statement that its partition function is the (q-th root of the) tau-function of 
the q-th Gelfand-Dickey hierarchy, which satisfies the string equation, {\it i.e.,} the  $L_{-1}$ Virasoro constraint.

Similarly to the KdV case, the boundary partition functions  are obtained as the exponential
vertex operators constructed from the bosons $\phi_l(x)$,
\be \label{openpf}
Z_{open}(t,z) = \langle \exp (\Gamma
\sum_{\alpha=1}^{q-1} \phi_{\alpha}(z)) \rangle\;,
\ee
where $\Gamma$ is a boundary component number counting parameter.

The open correlation numbers are then given by
\be
\langle \OO_{\alpha_1,k_1} \cdot \ldots \cdot \OO_{\alpha_n, k_n} \rangle_{open} :=
\frac{\pd}{\pd t_{\alpha_1, k_1}} \ldots \frac{\pd}{\pd t_{\alpha_n, k_n}}
\log (Z_{open}(t,z)/Z(t))\;.
\ee
From this expression one gets~\cite{Johnson:1993vk} the open $\cW$-constraints as the Ward identities for the
vertex operator
\be
L^{full}_n = L_n - \frac{\Gamma^2}{2} (n+1)z^n - z^{n+1} \frac{\pd}{\pd z}\;.
\ee
Let us take a formal Laplace transform of this operator in the variable $z$:
\be
\hat L^{full}_n = L_n - \frac{\Gamma^2}{2} (n+1)\frac{\pd^n}{\pd s^n} 
- s \frac{\pd^{n+1}}{\pd s^{n+1}}\;.
\ee
Having in mind the interpretation of $Z_{open}$ and $Z$ as disconnected surfaces
partition functions, or tau-functions, let us now represent
\be
Z_{open}(t,s) = \exp(F_c + F_o), \;\; F_c = \log(Z(t))\;.
\ee
Using the ordinary string equations~\eqref{strdil} for $Z(t)$, the
open string equation and the dilation equations for $F_o$ correspondingly read:
\ba \label{openstringdil}
&\sum_{\alpha,k} (\alpha/q+k) t_{\alpha,k}
\frac{\pd}{\pd t_{\alpha,k-1}} F^o = s \;,\\
&\sum_{\alpha,k} (\alpha/q+k) t_{\alpha,k}
\frac{\pd}{\pd t_{\alpha,k}} F^o = \Gamma^2/2 + s\frac{\pd}{\pd s} F^o\;.
\ea

After some renormalization these equations coincide with the equations obtained 
recently in~\cite{Buryak:2018ypm} for the generating function $F^{1/q,o}$ of the open
correlation numbers in the topological gravity of $q$-spin curves.\footnote{We note that in the same paper
the authors give an expression for $F^{1/q,o}$ in terms of the wave function
of the KP hierarchy.}

\paragraph{Loop operator.}

The analogue of the formula for the loop operator is easily obtained from the
bosonic representation~\cite{Johnson:1993vk} together with the Laplace transform:
\ba \label{loopq}
&w_r(l) = \sum_{k=0}^{\infty} \frac{l^{k+r/q}}{\Gamma(k+r/q+1)} \OO_{r,k}\;, \\
&w_r(l) \sim \int_{l} \frac{\dd l}l \, e^{-lz} \langle \phi_r(z) \rangle\;.
\ea

The loop insertion is the one-boundary part of the open partition 
function~\eqref{openpf}. In this case there are in general $q-1$ linearly independent loop operators. They can be 
interpreted as corresponding to different boundary conditions. However, the precise identification
with the FZZT branes is yet to be clarified. The general loop operator can be written as
\be \label{loopq1}
w(l):=\sum_{\alpha=1}^{q-1} c_{\alpha} w_{\alpha}(l)\;.
\ee
Below we omit the coefficients $c_{\alpha}$ as it will be
trivial to restore them in the final answer.

The analogs of open-Virasoro and $\cW$-constraints are then obtained from the Ward identities
for the fields $\phi_r(z)$, for instance:
\be
\langle T(x) \phi_r(z) \rangle \sim \frac{\pd \phi_r(z)}{z-x}\;,
\ee
from which we get 
\be
L^{loop}_n \, \langle \phi_r(z)\rangle =\left(L_n-z^{n+1}\frac{\pd}{\pd z}\right) \, \langle \phi_r(z)\rangle = 0, \;\; n\ge-1\;,\\
\ee
where $L_n$ are the closed Gelfand-Dickey Virasoro constraints.

Let us denote by $\hat \OO_{\alpha,k}$ the  bulk insertion operator
$\pd/\pd \lambda_{\alpha,k}$ in the MLG frame.
Then using the resonance transformations,
that is change of the couplings $t_{\alpha,k} \to \lambda_{\alpha, k}$ from
KdV to MLG frame, we obtain
\be \label{1ptpq}
\langle \hat \OO_{\alpha, k} \sum_r w_r(l) \rangle = u^{-k-\alpha/q} 
I_{k+\alpha/q}(2lu)\;,
\ee
which is precisely the singular part of the Liouville one-point boundary (FZZ) 
function found in~\cite{Fateev:2000ik}. This is one of the main results of the present paper.
Moreover, if we include the regular part of bosons in the definition
of the loop, we correctly restore also the regular part of FZZ formula.
The important point is that the resonance transformations have been computed from the
condition of diagonality of two-point functions in the MLG frame, corresponding to the Liouville couplings 
$\lambda_{\alpha,k}$, in the spherical topology.

Let us sketch the derivation of the formula~\eqref{1ptpq}. The detailed computation is 
given in Appendix~\bref{sec:appendix}.
First, we expand the sum $\sum_r w_r(l)$ as
\be
\sum_{\alpha=1}^{q-1}\sum_{k\ge 0} \frac{l^{\alpha/q+k+1}}{\Gamma(\alpha/q+k+1)}
\frac{\pd}{\pd t_{\alpha,k}}\;.
\ee
Then we express the left hand side  of~\eqref{1ptpq}  in terms of the two point functions
\be
\sum_{\beta, m}\frac{l^{\alpha/q+k+1}}{\Gamma(\alpha/q+k+1)}
\langle \hat \OO_{\alpha, k} \OO_{\beta, m} \rangle
\ee
and compute this expression using the results of~\cite{Belavin:2013nba} and \cite{Belavin:2014cua}, namely the formula for the generating function of the correlators and the explicit expression for the resonance transformation.

We shall now briefly describe the formulation of the method   of computing the correlation
numbers, based on the Frobenius manifold structure, leading to the results~\eqref{1ptpq}.

\subsection{Dual approach and Frobenius manifolds}

Here we formulate the result for the spherical partition function, corresponding to 
the tau-function of the $q-$th Gelfand-Dickey integrable hierarchy. The $L_{-1}$ constraint
(which uniquely fixes the tau-function) is written in the form 
of  Douglas
string equation, conveniently formulated as the action
principle~\cite{Ginsparg:1990zc},
$ \partial S(u) / \partial u_i= 0$,
and 
\be\label{actionInt}
S(u,t) = \Res_{y=\infty} \sum_{\alpha=1}^{q-1}\sum_{k=0}^{\infty}
t_{\alpha,k} Q^{\frac{\alpha}{q}+k}\;.
\ee
Here $Q = Q(y,u)$ is the symbol of the Lax operator of the corresponding
$q-$th Gelfand-Dickey hierarchy,
\be\label{Q-pol}
Q(y,u)=y^q+u_1 y^{q-2}+u_2 y^{q-3}+...+u_{q-1}\;.
\ee
The parameters $u_\alpha$, ($\alpha=1,...,q-1$), can be regarded as coordinates on the Frobenius manifold $A_{q-1}$. It's tangent space at a point $u$ is a Frobenius
algebra $\mathbb{C}[y] \mod \frac{\partial Q(y,u)}{\partial y}$ (for more details, see, e.g.,~\cite{Belavin:2014hsa}).
We note that for the $(q,p)-$model  the time parameter in~\eqref{actionInt} in front of $Q^{p/q}$
is equal to $1$ and the time parameter in front of $Q^{p/q-1}$ is equal to the Liouville cosmological constant $\mu$.

In order to consider the general MLG$(q,p)$ case it is convenient \cite{Belavin:2014hsa} to
use another parametrization,  $s$ and $p_0$, such that $p = sq +p_0$ and
$0 < p_0 < q$. As described in Sec.~\bref{section:prelim}, the physical fields $\OO_{m,n}$ are labeled by pairs $(m,n)$, where $1\geq m \geq q-1$ and $1\geq m \geq q-1$. 
Equivalently, in the ``KdV frame'', we use the parameters $(\alpha,k)$:
\be 
\alpha =p_0 m \text{ mod }q\,, \qquad k=s m-n + [p_0 m/q]\;.
\ee
The action can be rewritten as
\be\label{eq:pre-S}
 S(u, t[\lambda]) =  \text{Res}_{y = \infty} \left(Q^{\frac{p+q}{q}} + \!\!\sum_{(m,n)\in Kac(q,p)} \!\!t^{(m, n)} Q^{\frac{|pm-qn|}{q}} \right)\;.
\ee

The KdV times $t^{(m,n)}$ and  the Liouville couplings $\lambda_{mn}$ are related through the resonance transformation,  
\begin{equation}\label{resonance-transform}
t^{(m, n)} = \lambda_{m, n} + \sum A_{(m_1, n_1),(m_2, n_2)}^{(m, n)} \lambda_{m_1, n_1} 
\lambda_{m_2, n_2} + \cdots\;,
\end{equation} 
with the coefficients $A_{(m_1, n_1),(m_2, n_2)}^{(m, n)}$ constraint by the scaling properties and fixed by the underlying CFT selection rules~\cite{Belavin:2014hsa}.

We will perform the computations in the $t_{\alpha,k}$ frame, as it makes the
connection with the integrable structure more transparent.
Following \cite{Belavin:2014hsa}, we define
\be
\theta_{\alpha}(z):=\sum_{k\ge0} \theta_{\alpha,k}z^k\;,
\ee
where
\begin{equation}
\theta_{\alpha,k}=-c_{\alpha,k} \underset{y=\infty}{\text{res}} Q^{k+\frac{\alpha}{q}}(y)
\label{theta}
\end{equation}
and
\be
c_{\alpha,k}=\frac{\Gamma(\frac{\alpha}{q})}{\Gamma(\frac{\alpha}{q}+k+1)}\;.
\ee
Then the action takes the form
\be
S(u, t[\lambda]) =-\bigg[ \frac{\theta_{p_0,s}}{c_{p_0,s}} +\sum_{\sigma,k} t_{\sigma,k} \frac{\theta_{\sigma,k}}{c_{\sigma,k}} \bigg]\;,
\ee 
where $t(\lambda)$ stands for the resonance transformation \eqref{resonance-transform}. The generating function of the spherical correlators  is
\be \label{genfrob}
Z[t(\lambda)]=\int_0^{v^1_*} dv^{1} C^{\beta\gamma}_1 \frac{\partial S}{\partial v^\beta}\frac{\partial S}{\partial v^\gamma}\;,
\ee
where $v_\alpha$ and $C^{\beta\gamma}_\sigma$ are correspondingly flat
coordinates and structure constants of the Frobenius manifold
$A_{q-1}$, and $v_* \in A_{q-1}$ is the special solution of the string action.

\section{Resolvent expectation value, heat kernel problem and mirror symmetry}
\label{section:previous}

In this section we compare the loop formulae~\eqref{loopq},~\eqref{loopq1} with the results 
in~\cite{Aleshkin:2017yty}, where a different way 
to compute open MLG correlators was chosen. Namely, the following
expression for the loop operator in the $(q,p)$ model has been proposed there:
\be \label{looposc}
\tilde{w}(l) := \int_{t_{1,0}}^{\infty}
\dd x \int_{\gamma} \dd y \, e^{lQ(y, v(x))}\;.
\ee
Here $Q(y, v)$ is defined in~\eqref{Q-pol} and $v(x)$ is the solution of the string equation, corresponding to the closed string partition function. 
Let us compare the formula~\eqref{looposc} with the formula~\eqref{loopq}, and recall the difficulties encountered
in~\cite{Aleshkin:2017yty}.

First of all it is more convenient to write the derivative of the normalized loop
\be \label{loopder}
\pd_{t_{1,0}} \sum_{\beta} c_{\beta} w_{\beta}(l) = \sum_{\beta,m}
c_{\beta} \frac{l^{\beta/q+m}}{\Gamma(\beta/q+m+1)}
\pd_{t_{1,0}}\pd_{t_{\beta,m}} \log Z(t)\;,
\ee
where $Z(t)$ is the spherical partition function.
It is known, see, e.g.,~\cite{Belavin:2013nba}, that the second derivative $\pd_{t_{1,0}}\pd_{t_{\beta,m}} Z(t)
= \Res_{y=\infty} \, Q(y)^{\alpha/q+k}$.\footnote{Here we do not have a prefactor
as in~\cite{Belavin:2013nba} due to the different normalization of the action
$S = \sum_{\alpha,k} t_{\alpha,k} \Res\, Q(y)^{\alpha/q+k}$.}

To analyze the difference with the earlier approach~\cite{Aleshkin:2017yty},  we compare~\eqref{loopder} with the derivative 
of~\eqref{looposc}:
\be\label{2form}
\int_{\gamma} \dd y \, e^{lQ(y, v(x))} \;\;  \overset{?}{=}\;\;
\sum_{\beta=1}^{q-1} c_{\beta} \sum_{m\ge0} \frac{l^{\beta/q+m}}
{\Gamma(\beta/q+m+1)} \Res_{y=\infty} \, Q(y)^{\beta/q+m}\;.
\ee

We note, that if satisfied the relation~\eqref{2form} would reflect the classical genus zero mirror symmetry.
Indeed, according to the extended Witten's 
conjecture, the RHS is an analytic continuation of a
certain power series, counting the intersection numbers
on the moduli space of curves with $q-$spin structure, namely an $A$-model
expression. Whereas the left hand
side is a period integral for the dual $B$-model, which is an oscillating integral
of a Landau-Ginzburg model $W(y) = Q(y)$, or an $A_{q-1}$ singularity. In our
case we can simply establish the explicit connection. From this point of view,
the genus zero loop is a period of the mirror model with the deformation 
parameters $v_i$, as functions of couplings $t$ governed by the string equation.

We note, that for $q=2,$ that is in the KdV case, the equality above
is known as an asymptotic expansion of the heat kernel operator,
where the residues in the RHS are dispersionless analogues of the Seeley 
coefficients~\cite{DZ1}. A very similar phenomenon occurs in the arbitrary $q$
case.

Let us first expand the left hand side of the equation~\eqref{2form}. 
We introduce a notation
$Q(y) = y^q+Q_0(y),$ where $Q_0(y)$ is of degree $q-2$ in $y$. Then
\ba \label{oscformula}
&\int_{\gamma} \dd y \, e^{lQ(y, v(x))} = \int_{\gamma}\dd y \,
e^{ly^q} \sum_n \frac{l^n Q_0(y)^n}{n!} = \\ = &
\sum_{\alpha=0}^{q-2}\sum_{k,n\ge0} c_{\alpha}^{\gamma} e^{\pi i k}\frac{[y^{\alpha+kq}]
Q_0(y)^n}{n!}
\Gamma\left(\frac{\alpha+1}q +k\right) l^{n-k-\frac{\alpha+1}{q}}\;,
\ea
where we introduced a notation $[y^n](\sum p_k y^k) := p_n$ and
in the second line we computed the integral term by term in powers 
$\alpha+kq$ of $y$ as 
\be
\int_{\gamma} \dd y \, e^{y^q} y^{\alpha +kq} = 
c_{\alpha}^{\gamma} \Gamma\left(\frac{\alpha+1}{q}+k\right)\;,
\ee
where $c_{\alpha}^{\gamma}$ are some (in general complex) coefficients
of the expansion of the cycle $\gamma$ in a certain basis 
$(\Gamma_{\alpha})_{\alpha}$ in homology $H_1(\CC, \Re y^q << 0; \CC)$.
This basis is defined by duality
\be
\int_{\Gamma_{\alpha}} e^{-y^q} \, y^{\beta} = \delta_{\alpha,\beta}\;, \;\;
\beta \in [0,q-2]\;.
\ee

Now we turn to the right hand side of~\eqref{2form},
\be
\Res \, Q(y)^{\beta/q+m} = \sum_n [y^{-\beta-1+(n-m)q}]Q_0(y)^n 
\frac{\Gamma(\beta/q+m+1)}{\Gamma(\beta/q+m+1-n) \, n!}\;.
\ee
Then the right hand side of the equation~\eqref{2form} becomes
\be
\sum_{\beta} c_{\beta} \sum_{n,m} \frac{[y^{-\beta-1+(n-m)q}]Q_0(y)^n}{n!} 
\Gamma(\beta/q+m+1-n)^{-1} l^{\beta/q+m}\;.
\ee
Using reflection relation for gamma function and changing summation variables we
get
\be \label{resformula}
\sum_{\beta} c_{\beta} \sum_{k\ge0, \, n\ge 0} \frac{[y^{\alpha+kq}]Q_0(y)^n}{n!} 
\Gamma\left(\frac{\alpha+1}q+k\right) \frac{\sin\pi((\alpha+1)/q+k)}{\pi}
l^{n-k-\frac{\alpha+1}q}\;.
\ee

Now it is clear, that the formulae~\eqref{oscformula} and~\eqref{resformula}
differ by some constants and coefficients $c_{\beta}$ and 
$c_{\alpha}^{\gamma}$. Basically, $c_{\alpha}^{\gamma}$ in the expression~\eqref{looposc} is defined
by the cycle $\gamma,$ whereas in the approach of the present paper it is a matter
of choice of a particular linear combination $w(l):=\sum_{\beta} c_{\beta}
\, w_{\beta}(l)$. It is tempting to interpret $c_{\beta}$ as a boundary
condition of the minimal model, however it requires further investigation. From this point of view the vanishing of one-point correlators
of the form $\langle \hat \OO_{2\alpha, m} \rangle^{\text{disk}}$ for even $q$, encountered in~\cite{Aleshkin:2017yty},  is
just  related to the fact that the corresponding cycle $\gamma$ does not contain
some of $\Gamma_{\alpha}$ in the expansion over this basis. Another problem encountered in~\cite{Aleshkin:2017yty} was the problem in the computation of one-point functions in the non-unitary $(q,p)$ models (i.e. $p>q+1$), which was due to
inappropriate solution of the string equation in $\dd x$ integration.

\section{Concluding remarks}
 \label{section:concl}

 This paper is another step to make correspondence between Minimal Liouville
 Gravity and integrable hierarchies and topological string.
Here we make manifest the correspondence between integrable 
systems/topological gravity and CFT approach to Minimal Gravity on the level
of disk correlation numbers. Compared to the previous paper~\cite{Aleshkin:2017yty}
the method proposed here is more transparent and easily connected to the
integrable systems. It is formulated most easily in the language of twisted
free bosons. The approach used here also solves problems, which remained unclear in
the previous approach, motivated by multi-matrix models~\cite{Aleshkin:2017yty}. The connection
of the two formulas for the loop operators is given by a mirror symmetry type
formula.

We performed checks for arbitrary $(q,p)$-models using resonance transformations,
which was not yet done before. Instead of the matrix model approach we use Gelfand-Dickey integrable systems to compare with
the Liouville Gravity because it highlights the symmetries and seems closer to
the enumerative geometry and topological gravity. This approach also shows how
differential constraints appear in the minimal Liouville Gravity after resonance
transformations.

The computations are similar for any $(q,p)$ and use already known results on
spherical correlators and resonance transformations. The latter ones give
correspondence with FZZ formulae for disk one point correlation numbers. Different
choices of the loop operator as linear combination of different twisted bosons
give different boundary conditions, which explains some problems encountered 
in~\cite{Aleshkin:2017yty}, however the exact identification of FZZT branes
 requires further investigation.

\paragraph{Acknowledgements}
The authors are grateful to A. Alexandrov, B. Dubrovin and
G. Ruzzo for valuable discussions.

\appendix

\section{Computation of one-point correlation numbers} 
\label{sec:appendix}

As explained in Section~\ref{section:general}, the macroscopic loop $w(l)$, where $l$ is
the length of the loop, is created by the operator
\be \label{loop1}
w(l) =\sum_{\beta,j} \frac{l^{\beta/q+j}}{\Gamma(\beta/q+j+1)} \frac{\partial}{\partial t_{\beta,j}}\;,
\ee
where $t_{\beta,j}$ are KdV times. 
The one-point function of the bulk operator $\hat \OO_{\alpha,k}$  on a disk 
\be\label{O-disk-1}
\langle \hat \OO_{\alpha,k}\rangle^{\text{disk}} = \langle \hat \OO_{\alpha,k} \cdot w(l)  \rangle^{\text{sphere}}
\ee
is obtained from the generating function \eqref{genfrob} as follows
\be\label{O-disk-2}
  \langle \hat \OO_{\alpha,k} w(l) \rangle^{\text{sphere}}=\sum_{\beta,j} \frac{l^{\beta/q+j}}{\Gamma(\beta/q+j+1)} 
  \frac{\partial}{\partial t_{\beta,j}}\cdot\frac{\partial}{\partial \lambda_{\alpha,k}} \log Z[t(\lambda)]\;.
\ee
Note that the second derivative is taken with respect to $\lambda_{\alpha,k}$, since we are interested in the correlation functions
in the Liouville frame. Here comes non-trivial dependence on the resonance transformations \eqref{resonance-transform}.

Using \eqref{O-disk-1},\eqref{O-disk-2}, one gets
\be
  \langle \OO_{m,n}  \rangle^{\text{disk}}=\sum_{\beta,j} \frac{l^{\beta/q+j}}{\Gamma(\beta/q+j+1)} 
 \int_0^{v_*} dv^{\sigma} C^{\beta\gamma}_\sigma \left(-\frac{1}{c_{\beta,j}}\right)\frac{\partial \theta_{\beta,j}}{\partial v^\beta}
 \frac{\partial \hat S_{\alpha,k}}{\partial v^\gamma}\;.
\ee

It is convenient to take the integration contour\footnote{This is possible due to specific  properties of the integral representation and of the special solution $v_*$ of the string equation, for more details, see~\cite{Belavin:2014cua}.}  along $v_1$-axis and to use the properties of  the derivatives  $\frac{\partial S^{(m,n)}}{\partial v^\gamma}$ and of the structure constants on the line $v_1$, obtained in \cite{Belavin:2014cua}. 
Namely, using expressions for structure constants, one gets
\be\label{one-pt}
  \langle \hat \OO_{\alpha,k}  \rangle^{\text{disk}}=\sum_{\beta,j} \frac{l^{\beta/q+j}}{\Gamma(\beta/q+j+1)} 
\left(-\frac{1}{c_{\beta,j}}\right) \sum_{\gamma=1}^{q-1}\int_0^{v_1^0} d v_1 \left(-\frac{v_1}{q}\right)^{\gamma-1} 
 \frac{\partial \theta_{\beta,j}}{\partial v_\gamma}
 \frac{\partial \hat S_{\alpha, k}}{\partial v_\gamma}\;.
\ee

Because the expressions of $\hat S_{\alpha, k}$ and $\theta_{\beta,j}$
differ for odd and even $k,j$ we consider four computations separately.

\paragraph{First case.} 

Here we compute the correlation function for a field with even $k$.
\ba \label{evencorr1}
&\sum_{\beta,j} \frac{l^{\beta/q+j+1}}{\Gamma(\beta/q+j+1)}
\langle \hat \OO_{\alpha, 2k} \cdot  \OO_{\beta, j}\rangle = \\ =
\sum_{\beta,m} \frac{l^{\beta/q+2m+1}}{\Gamma(\beta/q+2m+1)}
&\langle \hat \OO_{\alpha, 2k} \cdot  \OO_{\beta, 2m}\rangle +
\sum_{\beta,m} \frac{l^{\beta/q+2m+2}}{\Gamma(\beta/q+2m+2)}
\langle \hat \OO_{\alpha, 2k} \cdot  \OO_{\beta, 2m+1}\rangle\;.
\ea

To compute the first summand in~\eqref{evencorr1} we expand it using~\eqref{one-pt}
\be \label{even1}
\langle  \hat \OO_{\alpha, 2k} \cdot  \OO_{\beta, 2m} \rangle = 
\sum_{\gamma=1}^{q-1}  \int\limits_0^{v_1^*} \dd (-v_1/q)^{\gamma} \, \frac{\pd
S_{\beta,2m}}{\pd v_{\gamma}} \frac{\pd \hat S_{\alpha, 2k}}{\pd v_{\gamma}}\;.
\ee

We use expressions from~\cite{Belavin:2014hsa}:
\ba \label{res1}
&\frac{\pd S_{\beta,2m}}{\pd v_{\gamma}} = -\delta_{\beta, \gamma}/c_{\beta, 2m}
\frac{\pd \theta_{\beta,2m}}{\pd v_{\gamma}} = \frac{\Gamma(\alpha/q+2m+1)}{
    \Gamma(\alpha/q+m)m!}x^m\;, \\
&\frac{\pd \hat S_{\alpha,2k}}{\pd v_{\gamma}} = \delta_{\alpha,\gamma} \, x_0^k
\,
P_k^{(0,\alpha/q-1)}(2x/x_0-1)\;,
\ea
where $x:= (-v_1/q)^q$ and $x_0 := (-v^*_1/q)^q$,
and explicit formula for Jacobi polynomial:
\be
P^{(0,\beta)}_k(2z-1)=\frac{1}{k!} z^{-\beta} \pd_z \left[ z^{\beta+k}(1-z)^k \right]\;.
\ee
When we plug all this expressions into~\eqref{even1}, we get:
\be \label{even2}
  \frac{\Gamma(\alpha/q+2m+1)}{
      \Gamma(\alpha/q+m)m!}\frac{x_0^{m+k+\alpha/q}}{k!} \int_0^{x_0} \dd (x/x_0) \, (x/x_0)^{m} 
 \pd_{\frac{x}{x_0}} \left[ (x/x_0)^{\alpha/q+k} \left(1-(x/x_0)^k\right) \right]\;.
\ee

After using Leibniz rule $k$ times, the last integral becomes beta function integral
\be
\int_{0}^1 \dd x \pd_x \left[ x^{\beta+k} (x-1)^k \right] = \frac{k!(m-k+1)_k}
{(\beta+m)_{k+1}}\;.
\ee

Inserting it into the formula~\eqref{even2} we obtain
\be
\langle  \hat \OO_{\alpha, 2k} \cdot  \OO_{\beta, 2m} \rangle = 
\frac{\Gamma(\alpha/q+2m+1) x_0^{m+k+\alpha/q}}{\Gamma(\alpha/q+m+k+1)\, (m-k)!}\;.
\ee
Finally, summing over $m$ with weight $l^{\alpha/q+2m}/\Gamma(\alpha/q+2m+1)$
and changing the summation variable $m\to m+k$ we get
\be \label{evenwitheven}
\langle\hat \OO_{\alpha,2k}\rangle^{\text{disk}} = (x_0^{1/2})^{\alpha/q+2k}
\sum_{m=0}^{\infty}\frac{(lx_0^{1/2})^{\alpha/q+2k+2m}}{\Gamma(\alpha/q+2k+m+1)\,m!}
=  (x_0^{1/2})^{\alpha/q+2k} I_{\alpha/q+2k}(2lx_0^{1/2})\;.
\ee

Now we compute the second summand from~\eqref{evencorr1}:
\be \label{even1}
\langle  \hat \OO_{\alpha, 2k} \cdot  \OO_{\beta, 2m+1} \rangle = 
\sum_{\gamma=1}^{q-1}  \int\limits_0^{v_1^*} \dd (-v_1/q)^{\gamma} \, \frac{\pd
S_{\beta,2m+1}}{\pd v_{\gamma}} \frac{\pd \hat S_{\alpha, 2k}}{\pd v_{\gamma}}\;,
\ee
where 
\be
 \frac{\pd
S_{\beta,2m+1}}{\pd v_{\gamma}} = -\delta_{q-\beta,\gamma} \,\frac{\Gamma(\alpha/q
+(2m+1)+1)}{\Gamma(\alpha/q+m+1)\, m!} x^{m+\alpha/q}\;.
\ee
Analogous to the previous case we get:
\be \label{even3}
  \frac{\Gamma(\alpha/q+(2m+1)+1)}{
      \Gamma(\alpha/q+m+1)\,m!}\frac{x_0^{m+k+1}}{k!} \int_0^{x_0} \dd (x/x_0) \, (x/x_0)^{m+1-\alpha/q} 
 \pd_{\frac{x}{x_0}} \left[ (x/x_0)^{\alpha/q+k} \left(1-(x/x_0)^k\right) \right]\;.
\ee
We notice, that this expression is analytic in $\mu$ and therefore should be 
disregarded in the expression as non-universal. However we proceed with computation
of this non-universal part because it gives interesting results.
Computing the integral above with same formula we obtain:
\be
\langle  \hat \OO_{\alpha, 2k} \cdot  \OO_{\beta, 2m+1} \rangle = 
-\frac{\Gamma(1-\alpha/q+2m+2)\, x_0^{m+k+1}}{\Gamma(-\alpha/q+m-k+1) \, (m+k+1)!}\;.
\ee

Finally we sum over $m$ and perform a variable shift $m\to m-k-1$:
\be \label{evenwithodd}
\sum_{m=k+1}^{\infty} -(x_0^{1/2})^{-\alpha/q-2k-1}\frac{(lx_0^{1/2})^{-\alpha/q-2k-1+2m}}
{\Gamma(-\alpha/q-2k-1+2m+1)\, m!}\;.
\ee
We note, that if in the definition of the loop~\eqref{loop1} we add regular terms,
that is to consider the summation range from $-\infty$ to $\infty$ treating 
differentiation wrt negative times as multilication by conjugated time, or as 
a pseudodifferential equation, then the result in~\eqref{evenwitheven} will not
change whereas the formula~\eqref{evenwithodd} the summation will be from $0$ to
$\infty$, yielding Bessel function
\be
\eqref{evenwithodd} = -(x_0^{1/2})^{\alpha/q+2k} \, I_{-\alpha/q-2k}(2lx_0^{1/2})\;.
\ee
When we add up both the contributions we get
\be
\langle \hat \OO_{\alpha,2k} \rangle^{\text{disk}} = \frac{2\sin(\alpha \pi)}{\pi} 
(x_0^{1/2})^{\alpha/q+2k}K_{\alpha/q}(2lx_0^{1/2})\;,
\ee
which coincides with the FZZ expression.

\paragraph{Second case.}

In this paragraph we compute
\be
\langle\hat \OO_{\alpha, 2k+1} \rangle^{\text{disk}} = \\ =
\sum_{\beta,m} \frac{l^{\beta/q+2m+1}}{\Gamma(\beta/q+2m+1)}
\langle \hat \OO_{\alpha, 2k+1} \cdot  \OO_{\beta, 2m}\rangle +
\sum_{\beta,m} \frac{l^{\beta/q+2m+2}}{\Gamma(\beta/q+2m+2)}
\langle \hat \OO_{\alpha, 2k+1} \cdot  \OO_{\beta, 2m+1}\rangle\;.
\ee
The situation is analogous to the even case: pairings with fields $\OO_{\beta,2m+1}$
are non-analytic and the ones with $\OO_{\beta,2m+1}$ are analytic.
First we compute the non-analytic part.
\be \label{odd1}
\langle  \hat \OO_{\alpha, 2k+1} \cdot  \OO_{\beta, 2m+1} \rangle = 
\sum_{\gamma=1}^{q-1}  \int\limits_0^{v_1^*} \dd (-v_1/q)^{\gamma} \, \frac{\pd
S_{\beta,2m+1}}{\pd v_{\gamma}} \frac{\pd \hat S_{\alpha, 2k+1}}{\pd v_{\gamma}}\;,
\ee
We again use the formula derived in~\cite{Belavin:2014hsa} from the diagonality condition for two
point functions:
\be \label{res2}
\frac{\pd \hat S_{\alpha, 2k+1}}{\pd v_{\beta}} = \delta_{\alpha,q-\beta}\,
x_0^k x^{\alpha/q} P_k^{0,\alpha/q}(2x/x_0-1)\;.
\ee
Formula~\eqref{odd1} becomes:
\be
\frac{-\Gamma(\alpha/q+2m+2)}{\Gamma(\alpha/q+m+1)\, m! \,k!}
\int_0^{x_0} x_0^{\alpha/q+m+1} \dd x  \, x^m \pd_x^k 
\left[ x^{\alpha/q+1+k}(1-x)^k \right]\;.
\ee
Non-analytic part the correlator becomes
\be
-(x_0^{1/2})^{\alpha/q+2k+1} \sum_{m=0}^{\infty} 
\frac{(lx_0^{1/2})^{\alpha/q+2k+1+2m}}{\Gamma(\alpha/q+2k+1+m+1)\,m!} = 
-(x_0^{1/2})^{\alpha/q+2k+1} I_{\alpha/q+2k+1}(2lx_0^{1/2})\;.
\ee

By the same argument, the nonanalytic part is equal to
\be
\sum_{m\ge0}\frac{\Gamma(1-\alpha/q+2m+1)}{\Gamma(1-\alpha/q+m)\, m!\,k!}
x_0^{m+k+1} \int_0^1 \dd x \, x^{m-\alpha/q} \pd_x^k \left[ 
x^{\alpha/q+k}(1-x)^k\right]\;,
\ee
which evaluates to
\be
(x_0^{1/2})^{\alpha/q+2k+1}\sum_{m\ge k+1} \frac{(lx_0)^{-\alpha/q-2k-1_2m}}
{\Gamma(m-\alpha/q-2k-1+1)\, m!}\;.
\ee

Similarly to the even case we see that up to a small mismatch in first $k+1$ terms
this coincides with the Bessel function. If we add regular terms to the 
loop definition, we get
\be
\langle \hat \OO_{\alpha,2k+1} \rangle^{\text{disk}} =\frac{2 \sin(\alpha\pi)}{\pi}
(x_0^{1/2})^{\alpha/q+2k+1} K_{\alpha/q+2k+1}(2lx_0^{1/2})\;.
\ee

\bibliographystyle{JHEP}
\bibliography{mlg2}

%\printbibliography

\end{document}